\documentclass[aps,prl,preprint,tightenlines,superscriptaddress,showpacs,byrevtex]{revtex4-1}

\usepackage{graphicx} 
\usepackage{dcolumn}  

\graphicspath{{ps}}

\newcommand{\Lc}{\Lambda_{c}^{+}}
\newcommand{\Sc}{\Sigma_{c}}

\newcommand{\Lcts}{\Lambda_{c}(2765)^{+}}
\newcommand{\Lcte}{\Lambda_{c}(2880)^{+}}
\newcommand{\Lctn}{\Lambda_{c}(2940)^{+}}
\newcommand{\pim}{\pi^{-}}
\newcommand{\pip}{\pi^{+}}

\RequirePackage{lineno}
\begin{document}

\preprint{\vbox{ \hbox{   }
                 \hbox{BELLE-CONF-1905}
}}

\title{ \quad\\[0.5cm] Experimental determination of the 
isospin of $\Lcts/\Sigma_c(2765)^+$ }

\noaffiliation

\noaffiliation
\affiliation{University of the Basque Country UPV/EHU, 48080 Bilbao}
\affiliation{Beihang University, Beijing 100191}
\affiliation{University of Bonn, 53115 Bonn}
\affiliation{Brookhaven National Laboratory, Upton, New York 11973}
\affiliation{Budker Institute of Nuclear Physics SB RAS, Novosibirsk 630090}
\affiliation{Faculty of Mathematics and Physics, Charles University, 121 16 Prague}
\affiliation{Chiba University, Chiba 263-8522}
\affiliation{Chonnam National University, Gwangju 61186}
\affiliation{University of Cincinnati, Cincinnati, Ohio 45221}
\affiliation{Deutsches Elektronen--Synchrotron, 22607 Hamburg}
\affiliation{Duke University, Durham, North Carolina 27708}
\affiliation{University of Florida, Gainesville, Florida 32611}
\affiliation{Department of Physics, Fu Jen Catholic University, Taipei 24205}
\affiliation{Key Laboratory of Nuclear Physics and Ion-beam Application (MOE) and Institute of Modern Physics, Fudan University, Shanghai 200443}
\affiliation{Justus-Liebig-Universit\"at Gie\ss{}en, 35392 Gie\ss{}en}
\affiliation{Gifu University, Gifu 501-1193}
\affiliation{II. Physikalisches Institut, Georg-August-Universit\"at G\"ottingen, 37073 G\"ottingen}
\affiliation{SOKENDAI (The Graduate University for Advanced Studies), Hayama 240-0193}
\affiliation{Gyeongsang National University, Jinju 52828}
\affiliation{Department of Physics and Institute of Natural Sciences, Hanyang University, Seoul 04763}
\affiliation{University of Hawaii, Honolulu, Hawaii 96822}
\affiliation{High Energy Accelerator Research Organization (KEK), Tsukuba 305-0801}
\affiliation{J-PARC Branch, KEK Theory Center, High Energy Accelerator Research Organization (KEK), Tsukuba 305-0801}
\affiliation{Forschungszentrum J\"{u}lich, 52425 J\"{u}lich}
\affiliation{Hiroshima Institute of Technology, Hiroshima 731-5193}
\affiliation{IKERBASQUE, Basque Foundation for Science, 48013 Bilbao}
\affiliation{University of Illinois at Urbana-Champaign, Urbana, Illinois 61801}
\affiliation{Indian Institute of Science Education and Research Mohali, SAS Nagar, 140306}
\affiliation{Indian Institute of Technology Bhubaneswar, Satya Nagar 751007}
\affiliation{Indian Institute of Technology Guwahati, Assam 781039}
\affiliation{Indian Institute of Technology Hyderabad, Telangana 502285}
\affiliation{Indian Institute of Technology Madras, Chennai 600036}
\affiliation{Indiana University, Bloomington, Indiana 47408}
\affiliation{Institute of High Energy Physics, Chinese Academy of Sciences, Beijing 100049}
\affiliation{Institute of High Energy Physics, Vienna 1050}
\affiliation{Institute for High Energy Physics, Protvino 142281}
\affiliation{Institute of Mathematical Sciences, Chennai 600113}
\affiliation{INFN - Sezione di Napoli, 80126 Napoli}
\affiliation{INFN - Sezione di Torino, 10125 Torino}
\affiliation{Advanced Science Research Center, Japan Atomic Energy Agency, Naka 319-1195}
\affiliation{J. Stefan Institute, 1000 Ljubljana}
\affiliation{Kanagawa University, Yokohama 221-8686}
\affiliation{Institut f\"ur Experimentelle Teilchenphysik, Karlsruher Institut f\"ur Technologie, 76131 Karlsruhe}
\affiliation{Kavli Institute for the Physics and Mathematics of the Universe (WPI), University of Tokyo, Kashiwa 277-8583}
\affiliation{Kennesaw State University, Kennesaw, Georgia 30144}
\affiliation{King Abdulaziz City for Science and Technology, Riyadh 11442}
\affiliation{Department of Physics, Faculty of Science, King Abdulaziz University, Jeddah 21589}
\affiliation{Kitasato University, Sagamihara 252-0373}
\affiliation{Korea Institute of Science and Technology Information, Daejeon 34141}
\affiliation{Korea University, Seoul 02841}
\affiliation{Kyoto University, Kyoto 606-8502}
\affiliation{Kyungpook National University, Daegu 41566}
\affiliation{LAL, Univ. Paris-Sud, CNRS/IN2P3, Universit\'{e} Paris-Saclay, Orsay 91898}
\affiliation{\'Ecole Polytechnique F\'ed\'erale de Lausanne (EPFL), Lausanne 1015}
\affiliation{P.N. Lebedev Physical Institute of the Russian Academy of Sciences, Moscow 119991}
\affiliation{Liaoning Normal University, Dalian 116029}
\affiliation{Faculty of Mathematics and Physics, University of Ljubljana, 1000 Ljubljana}
\affiliation{Ludwig Maximilians University, 80539 Munich}
\affiliation{Luther College, Decorah, Iowa 52101}
\affiliation{Malaviya National Institute of Technology Jaipur, Jaipur 302017}
\affiliation{University of Malaya, 50603 Kuala Lumpur}
\affiliation{University of Maribor, 2000 Maribor}
\affiliation{Max-Planck-Institut f\"ur Physik, 80805 M\"unchen}
\affiliation{School of Physics, University of Melbourne, Victoria 3010}
\affiliation{University of Mississippi, University, Mississippi 38677}
\affiliation{University of Miyazaki, Miyazaki 889-2192}
\affiliation{Moscow Physical Engineering Institute, Moscow 115409}
\affiliation{Moscow Institute of Physics and Technology, Moscow Region 141700}
\affiliation{Graduate School of Science, Nagoya University, Nagoya 464-8602}
\affiliation{Kobayashi-Maskawa Institute, Nagoya University, Nagoya 464-8602}
\affiliation{Universit\`{a} di Napoli Federico II, 80055 Napoli}
\affiliation{Nara University of Education, Nara 630-8528}
\affiliation{Nara Women's University, Nara 630-8506}
\affiliation{National Central University, Chung-li 32054}
\affiliation{National United University, Miao Li 36003}
\affiliation{Department of Physics, National Taiwan University, Taipei 10617}
\affiliation{H. Niewodniczanski Institute of Nuclear Physics, Krakow 31-342}
\affiliation{Nippon Dental University, Niigata 951-8580}
\affiliation{Niigata University, Niigata 950-2181}
\affiliation{University of Nova Gorica, 5000 Nova Gorica}
\affiliation{Novosibirsk State University, Novosibirsk 630090}
\affiliation{Osaka City University, Osaka 558-8585}
\affiliation{Osaka University, Osaka 565-0871}
\affiliation{Pacific Northwest National Laboratory, Richland, Washington 99352}
\affiliation{Panjab University, Chandigarh 160014}
\affiliation{Peking University, Beijing 100871}
\affiliation{University of Pittsburgh, Pittsburgh, Pennsylvania 15260}
\affiliation{Punjab Agricultural University, Ludhiana 141004}
\affiliation{Research Center for Electron Photon Science, Tohoku University, Sendai 980-8578}
\affiliation{Research Center for Nuclear Physics, Osaka University, Osaka 567-0047}
\affiliation{Theoretical Research Division, Nishina Center, RIKEN, Saitama 351-0198}
\affiliation{RIKEN BNL Research Center, Upton, New York 11973}
\affiliation{Saga University, Saga 840-8502}
\affiliation{University of Science and Technology of China, Hefei 230026}
\affiliation{Seoul National University, Seoul 08826}
\affiliation{Shinshu University, Nagano 390-8621}
\affiliation{Showa Pharmaceutical University, Tokyo 194-8543}
\affiliation{Soongsil University, Seoul 06978}
\affiliation{University of South Carolina, Columbia, South Carolina 29208}
\affiliation{Stefan Meyer Institute for Subatomic Physics, Vienna 1090}
\affiliation{Sungkyunkwan University, Suwon 16419}
\affiliation{School of Physics, University of Sydney, New South Wales 2006}
\affiliation{Department of Physics, Faculty of Science, University of Tabuk, Tabuk 71451}
\affiliation{Tata Institute of Fundamental Research, Mumbai 400005}
\affiliation{Excellence Cluster Universe, Technische Universit\"at M\"unchen, 85748 Garching}
\affiliation{Department of Physics, Technische Universit\"at M\"unchen, 85748 Garching}
\affiliation{Toho University, Funabashi 274-8510}
\affiliation{Tohoku Gakuin University, Tagajo 985-8537}
\affiliation{Department of Physics, Tohoku University, Sendai 980-8578}
\affiliation{Earthquake Research Institute, University of Tokyo, Tokyo 113-0032}
\affiliation{Department of Physics, University of Tokyo, Tokyo 113-0033}
\affiliation{Tokyo Institute of Technology, Tokyo 152-8550}
\affiliation{Tokyo Metropolitan University, Tokyo 192-0397}
\affiliation{Tokyo University of Agriculture and Technology, Tokyo 184-8588}
\affiliation{Utkal University, Bhubaneswar 751004}
\affiliation{Virginia Polytechnic Institute and State University, Blacksburg, Virginia 24061}
\affiliation{Wayne State University, Detroit, Michigan 48202}
\affiliation{Yamagata University, Yamagata 990-8560}
\affiliation{Yonsei University, Seoul 03722}
  \author{A.~Abdesselam}\affiliation{Department of Physics, Faculty of Science, University of Tabuk, Tabuk 71451} 
  \author{I.~Adachi}\affiliation{High Energy Accelerator Research Organization (KEK), Tsukuba 305-0801}\affiliation{SOKENDAI (The Graduate University for Advanced Studies), Hayama 240-0193} 
  \author{K.~Adamczyk}\affiliation{H. Niewodniczanski Institute of Nuclear Physics, Krakow 31-342} 
  \author{J.~K.~Ahn}\affiliation{Korea University, Seoul 02841} 
  \author{H.~Aihara}\affiliation{Department of Physics, University of Tokyo, Tokyo 113-0033} 
  \author{S.~Al~Said}\affiliation{Department of Physics, Faculty of Science, University of Tabuk, Tabuk 71451}\affiliation{Department of Physics, Faculty of Science, King Abdulaziz University, Jeddah 21589} 
  \author{K.~Arinstein}\affiliation{Budker Institute of Nuclear Physics SB RAS, Novosibirsk 630090}\affiliation{Novosibirsk State University, Novosibirsk 630090} 
  \author{Y.~Arita}\affiliation{Graduate School of Science, Nagoya University, Nagoya 464-8602} 
  \author{D.~M.~Asner}\affiliation{Brookhaven National Laboratory, Upton, New York 11973} 
  \author{H.~Atmacan}\affiliation{University of South Carolina, Columbia, South Carolina 29208} 
% ...
\collaboration{The Belle Collaboration}

\begin{abstract}%
   We report an experimental determination of the isospin of $\Lcts/\Sigma_c(2765)^+$
  using 980 fb$^{-1}$ data in the $e^+e^-$ annihilation 
  around $\sqrt{s} = 10.6$ GeV
  collected by the Belle detector located at the KEKB collider.
  The isospin partners are searched for in the 
  $\Sc(2455)^{++/0} \pi^{0}$ channels, and no evidence was obtained. Thus
 the isospin is determined to be zero,
and the particle is established to be a $\Lambda_c$.
\end{abstract}

\maketitle

{\renewcommand{\thefootnote}{\fnsymbol{footnote}}}
\setcounter{footnote}{0}

\section{Introduction}\label{intro}

Charmed baryons are an excellent system to study 
quark dynamics, especially possible diquark correlations in baryons,
as they consist of one heavy charm quark and two light quarks.
There are a few dozens of states known so far, but 
in most cases their quantum
numbers have not been determined experimentally \cite{PDG}.
For ground states and low-lying states with excitation energies
up to about 400 MeV, classification by quark models still
works well and plausible quantum numbers can be given. 
However, for higher excitation energies, unique identification 
is not possible because quark models predict
several states within their typical mass uncertainties of around 50 MeV/$c^2$
\cite{th_capstick, th-Roberts, th-yoshida}.

$\Lcts/\Sigma_c(2765)^+$ (for the notation, see note 
\cite{footnote-notation} in the reference section)
is the lightest charmed baryon for which
there is no assumed quark-model identification.
It was first observed by CLEO \cite{Lc27CLEO} in 
the $\Lc \pip \pim$ final state and
later reported by Belle \cite{Lc28Belle} 
in the $\Sc^{++/0} \pi^{\mp}$ final state.
The experimental status is very poor:
the spin, parity and isospin are not determined yet
and CLEO did not determine the uncertainty on its width.

In a quark model calculation by Capstick and Isgur \cite{th_capstick},
there are six states predicted in this mass region, one state
each for $I(J^P)=0(1/2^+),~0(1/2^-)$, and two states each for
$I(J^P)=1(1/2^-),~(3/2^-)$. On the other hand, 
in the most recent quark model
calculation by Yoshida {\em et al.} \cite{th-yoshida}, the $I=0$
states are found in the higher mass region, while the $I=1$ states are
still in the range of $2765\pm50$ MeV/$c^2$.
In addition, there are many theoretical calculations in various models
(for example \cite{th-oh,th-Zhong, th-Ebert, th-Roberts, th-chen, th-chen2,
th-kumakawa, th-cheng, th-Lu}), and almost any combination of
$I=0$ or 1, $J=1/2$ or 3/2, and $P=+$ or $-$ seems to be possible. 
Therefore, the nature of $\Lcts/\Sigma_c(2765)^+$
is quite controversial and experimental determination of its quantum
numbers is awaited. 

In this paper, we present a 
determination of the isospin of $\Lcts/\Sigma_c(2765)^+$ by 
searching for 
the possible isospin partners of the state in the 
decay modes to $\Sc(2455)^{++/0}\pi^0 \to
\Lc \pi^{\pm} \pi^0$.

\section{The Belle experiment and the data set}\label{Sect:dataset}

We use a dataset of 980 fb$^{-1}$ 
of integrated luminosity around $\sqrt{s} = 10.6$ GeV
collected by the Belle detector
at the KEKB asymmetric-energy $e^{+}e^{-}$ collider \cite{KEKB}.
The Belle detector is a large solid-angle magnetic
spectrometer that consists of a silicon vertex detector (SVD),
a 50-layer central drift chamber (CDC), an array of
aerogel threshold Cherenkov counters (ACC),  
a barrel-like arrangement of time-of-flight
scintillation counters (TOF), and an electromagnetic calorimeter
comprised of CsI(Tl) crystals (ECL) located inside 
a superconducting solenoid coil that provides a 1.5~T
magnetic field.  An iron flux-return located outside of
the coil is instrumented to detect $K_L^0$ mesons and to identify
muons. The detector is described in detail elsewhere~\cite{Belle}.
Two inner detector configurations were used. A 2.0 cm radius beam-pipe
and a 3-layer silicon vertex detector were used for the first sample
of 156 fb$^{-1}$, while a 1.5 cm radius beam pipe, a 4-layer
silicon detector and a small-cell inner drift chamber were used to record
the remaining 824 fb$^{-1}$ \cite{svd2}. 
 We use a Monte Carlo (MC) simulation to model the detector response
and its acceptance to obtain the reconstruction efficiency, 
the mass resolution for signals and feed-down signals.
 The MC events are generated with PYTHIA \cite{pythia},
decays of unstable particles are simulated with EVTGEN \cite{evtgen}, 
and the detector response is simulated by using GEANT \cite{geant}.


\section{Reconstruction of $\Lcts$ in $\Lc \pip \pim$}\label{Sect:reconstruction}

\begin{figure}[h!]
  \begin{center}
    \includegraphics[scale=0.45]{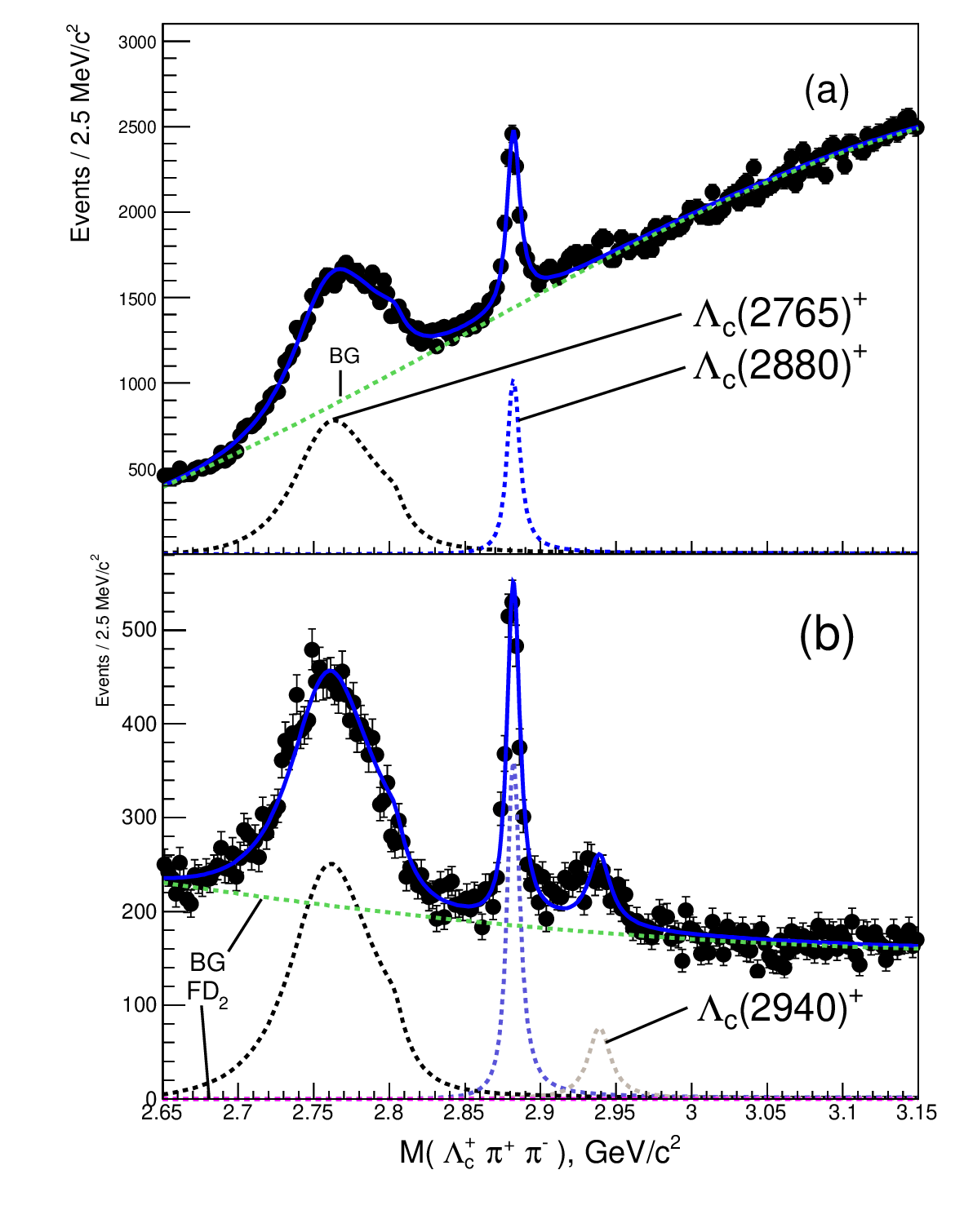}
    \caption{ 
      The $M(\Lc\pip\pim)$ distributions 
      with (a) no cut on $M(\Lc\pi)$ and 
      (b) $\Sc(2455)$ selection.
      The fit results (solid curves) and its  
      components (dashed curves) are also presented.
      Dashed black, blue, and gray lines are for the $\Lcts$, $\Lcte$,
      and $\Lctn$ signals, respectively, and the 
      dashed pink and green lines are for $\Sc(2520)$ feed-down (FD$_2$), 
      and the background (BG), respectively. 
    }
    \label{fig:lcpipi_1}
  \end{center}
\end{figure}

 We reconstruct $\Lcts$ candidates in the $\Lc \pip \pim$ final state.
 $\Lc$ baryons are reconstructed using the $\Lc \to p K^{-} \pi^{+}$ decay mode
(the inclusion of the charge-conjugate modes is implied throughout this paper).
For the primary selection of charged particles, 
we use the same track 
quality and particle identification 
as the $\Lcte$ study by Belle \cite{Lc28Belle}.
 In addition, in order to reject duplicated tracks, 
when there are multiple tracks of the same particle species
(pion, kaon, or proton), 
we check the invariant mass of each track pair. 
 If the invariant mass is smaller than 0.28 GeV/$c^{2}$, 
0.989 GeV/$c^{2}$, and 1.878 GeV/$c^{2}$
for pion, kaon, and proton pairs, respectively, 
one of the tracks is rejected by the track quality. 
 This selection reduces the number of 
$\Lc\pi^{+}\pi^{-}$ candidates by 0.1\%.
 The invariant mass of the $p K^{-} \pi^{+}$ combination 
is required to be within $\pm 8$ MeV/$c^{2}$ (3.7$\sigma$) 
of the mass of $\Lc$ \cite{ci_lc_w_babar}.
 We perform a $\Lc$-mass-constrained fit 
to the $p K^{-} \pi^{+}$ vertex to 
improve the $\Lc$ momentum resolution. 
 We combine the $\Lc$ candidate with the remaining $\pip\pim$ candidates.
 The scaled momentum $x_{p}=p^{*}/p_{\rm max}$,
 where $p^{*}$ is the momentum of the $\Lcts$ candidate 
in the beam center-of-mass frame, 
 $p_{\rm max}$ is given by $\sqrt{E_{\rm beam}^{2} - M^{2}c^4}$ 
with beam energy $E_{\rm beam}$ and mass of $\Lcts$ candidate $M$,
 is required to be greater than 0.8 to reduce combinatorial backgrounds. 
 To improve the invariant mass resolution,
 we perform a common vertex fit to each $\Lc \pip \pim$ candidate. 
 The resulting $\Lc \pip \pim$ invariant 
mass [$M(\Lc \pip \pim)$] spectrum 
is shown in Fig.~\ref{fig:lcpipi_1}(a), 
 which presents clear peaks of $\Lcts$ and $\Lcte$.

\begin{figure}[h!]
  \begin{center}
    \includegraphics[scale=0.4]{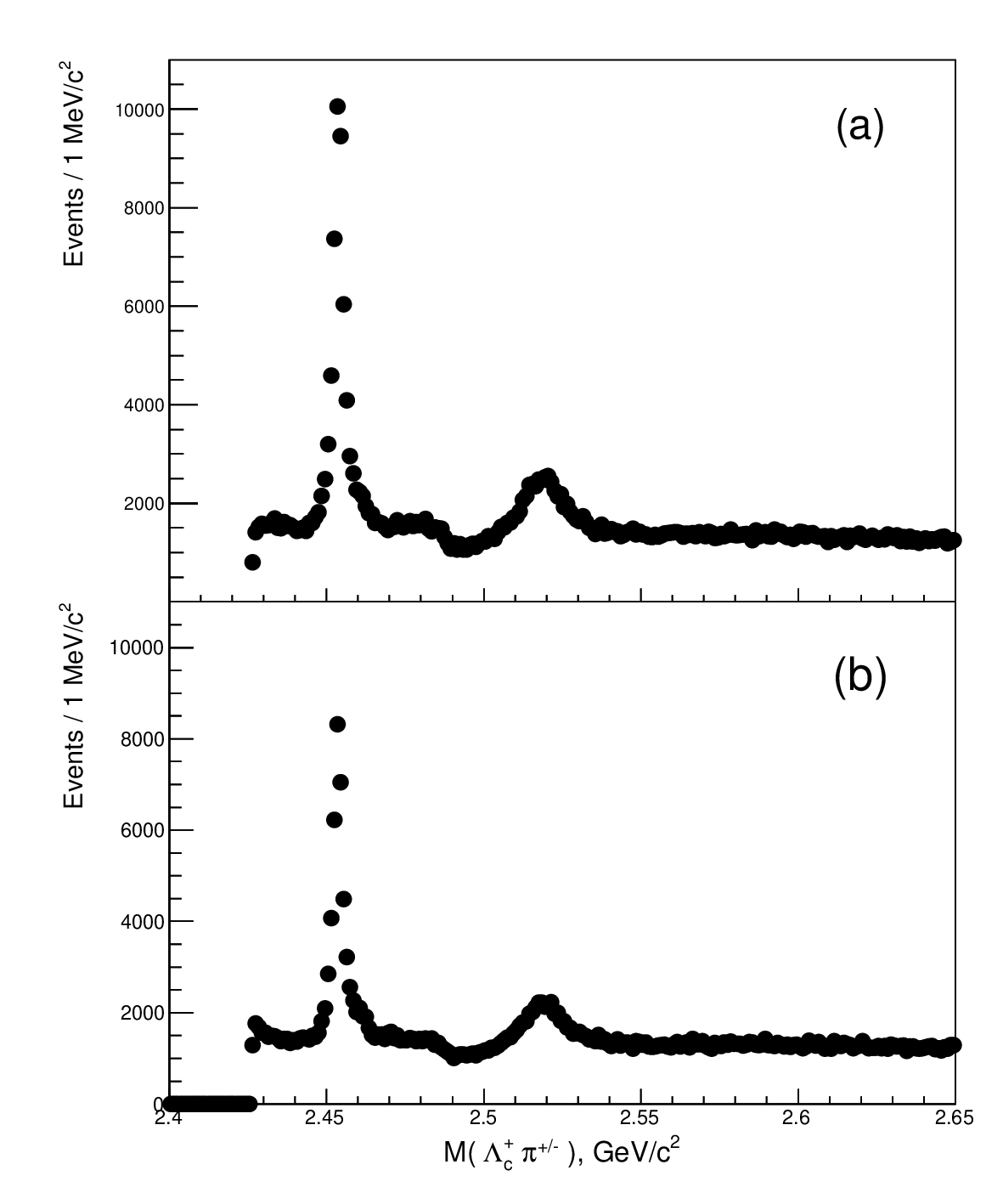} 
    \caption{ 
      The distributions of (a) $M(\Lc\pip)$ and (b) $M(\Lc\pim)$. 
    }
    \label{fig:lcpi_1}
  \end{center}
\end{figure}

 We can also see intermediate states, $\Sc(2455/2520)$,
 in the $M(\Lc\pi)$ distribution as shown in Fig.~\ref{fig:lcpi_1}.
 In the following analysis, we define the signal region for
 $\Sc(2455)$ as 
$\left|M(\Lc\pi)-m_{\Sc(2455)}\right| < 5$ MeV/$c^2$. 
Figure~\ref{fig:lcpipi_1}(b)
shows the $M(\Lc \pip \pim)$ spectrum with
$\Sc(2455)$ selection. 

The yield for $\Lcts \to \Sc(2455)^{++/0}\pi^{\mp} (\to \Lc \pip \pim)$ 
is determined by a fit to Fig.~\ref{fig:lcpipi_1}(b) using Breit-Wigner
functions for $\Lcts$, $\Lcte$, and $\Lctn$ and a second-order polynomial
as background. In addition, there is a small contribution from 
$\Sc(2520)$ feed-down which is due to an accidental coincidence 
of $\Sc(2520)$ and a random pion that produces a rather narrow peak 
in the $M(\Lc \pip \pim)$ spectrum around 2.68 GeV/$c^2$. 
The details are to be reported elsewhere \cite{Lc2765full}.

\section{Search for isospin partners in 
$\Lambda_{c}^{+}\pi^{\pm}\pi^{0}$ final state} 
\label{Sect:isospin}

In order to determine the isospin, we search for the isospin partners with 
$I_{3}=\pm1$ in the $\Sigma_c(2765)^{++/0} \to \Sigma_{c}^{++/0} \pi^{0} 
\to \Lambda_{c}^{+}\pi^{\pm}\pi^{0}$ final states.
The $\Sigma_c(2765)^{++/0} \to \Sigma_{c}^{+} \pi^{\pm} \to 
\Lambda_{c}^{+}\pi^{0}\pi^{\pm}$ decay modes are not used
because the momentum of $\pi^{0}$ in these modes is low and thus the background level is very high.
The ratio for the product of cross section and 
branching fractions between $I_{3}=0$ and $I_{3}=\pm 1$ states ($R_{I}$)
\begin{equation}\label{eq_r1}
	R_{I}=\frac{\sigma_{+1}{\cal B}_{+1} + \sigma_{-1}{\cal B}_{-1} }
                {2 \sigma_{0} {\cal B}_{0}},
\end{equation}
where $\sigma_{i}$ is the production cross section for the 
$I_{3}=i$ states, ${\cal B}_{\pm 1}$ is the 
branching fraction for the $I_{3}=\pm 1$ states 
decaying into $\Sigma_{c}(2455)^{++/0} \pi^{0}$ decay modes, and
${\cal B}_{0}$ is the sum of the branching fractions of the 
$I_{3}=0$ state decaying into $\Sigma_{c}(2455)^{++} \pi^{-}$
and $\Sigma_{c}(2455)^{0} \pi^{+}$, is measured.
If we assume the cross sections are the same for $I_{3}=+1$ and $I_{3}=-1$ and  
all the branching fractions are the same, this ratio simplifies to
\begin{equation}\label{eq_r2}
R_{I}=\frac{\sigma_{\pm1}} {\sigma_{0}}.
\end{equation}
If resonances are observed with masses and widths close to those of the
$I_{3}=0$ state,
it is strong evidence that the isospin is one.
If resonances are not observed and obtained upper limit for 
the $R_{I}$ is much smaller than one, it is strong evidence that the isospin is zero. 

In this analysis, the signal region, defined as $2.65$ GeV/$c^{2} < 
M(\Lambda_{c}^{+} \pi^{\pm} \pi^{0}) < 2.9$ GeV/$c^{2}$
after selecting the $\Sigma_{c}^{++/0}(2455)$ region, is blinded until 
all the selection criteria and combinatorial background are determined.
Except for the $\pi^{0}$, the event selection is the same 
as that applied for the $\Lambda_{c}^{+}\pi^{+}\pi^{-}$ analysis.
A $\pi^{0}$ is reconstructed from two photons,
energies of which are required to be greater than 50 MeV.
The $\pi^{0}$ candidates are selected by requiring 120 MeV/$c^{2}<M(2\gamma)<150$ MeV/$c^{2}$.
In order to suppress the combinatorial background further, the $\pi^{0}$ helicity angle,
defined as the angle between the momentum of the $\Lambda_{c}(2765)$ in the center-of-mass frame
and momentum of the $\pi^{0}$ in the $\Lambda_{c}(2765)^{+}$ rest frame is required to be 
greater than $-0.35$.

The mass distributions for combinatorial background events are 
checked from the $\Sigma_{c}(2455)^{++/0}$ 
sideband region defined as
$\left| M(\Lc \pi^{\pm})-m_{\Sigma_{c}(2455)^{++/0}} \pm 20 \text{ MeV}/c^{2}\right| < 5 \text{ MeV}/c^{2}$.
The $M(\Lc \pi^{\pm} \pi^{0})$ distribution for the sideband region is 
fitted with a second order exponential curve
$f(M) = \exp(aM+bM^{2})$, where $a$ and $b$ are the fit parameters.
The values of the $\chi^{2}/d.o.f.$ 
are obtained as $162.0/184$ for 
$M(\Lambda_{c}^{+} \pi^{+} \pi^{0})$ and 
212.2/184 for $M(\Lambda_{c}^{+} \pi^{-} \pi^{0})$,
confirming that this background parameterization is appropriate. 
Therefore it is decided to use this function for the background PDF.
The validity to use this function for the $\Sigma_{c}(2455)$ signal regions is 
checked by comparing mass distributions 
for sideband region and $\Sigma_{c}(2455)$ signal region 
outside of the signal box. They are found to be consistent with each other.

\subsection{Result}

\begin{figure*}[htbp]
  \begin{center}
    \includegraphics[scale=0.35]{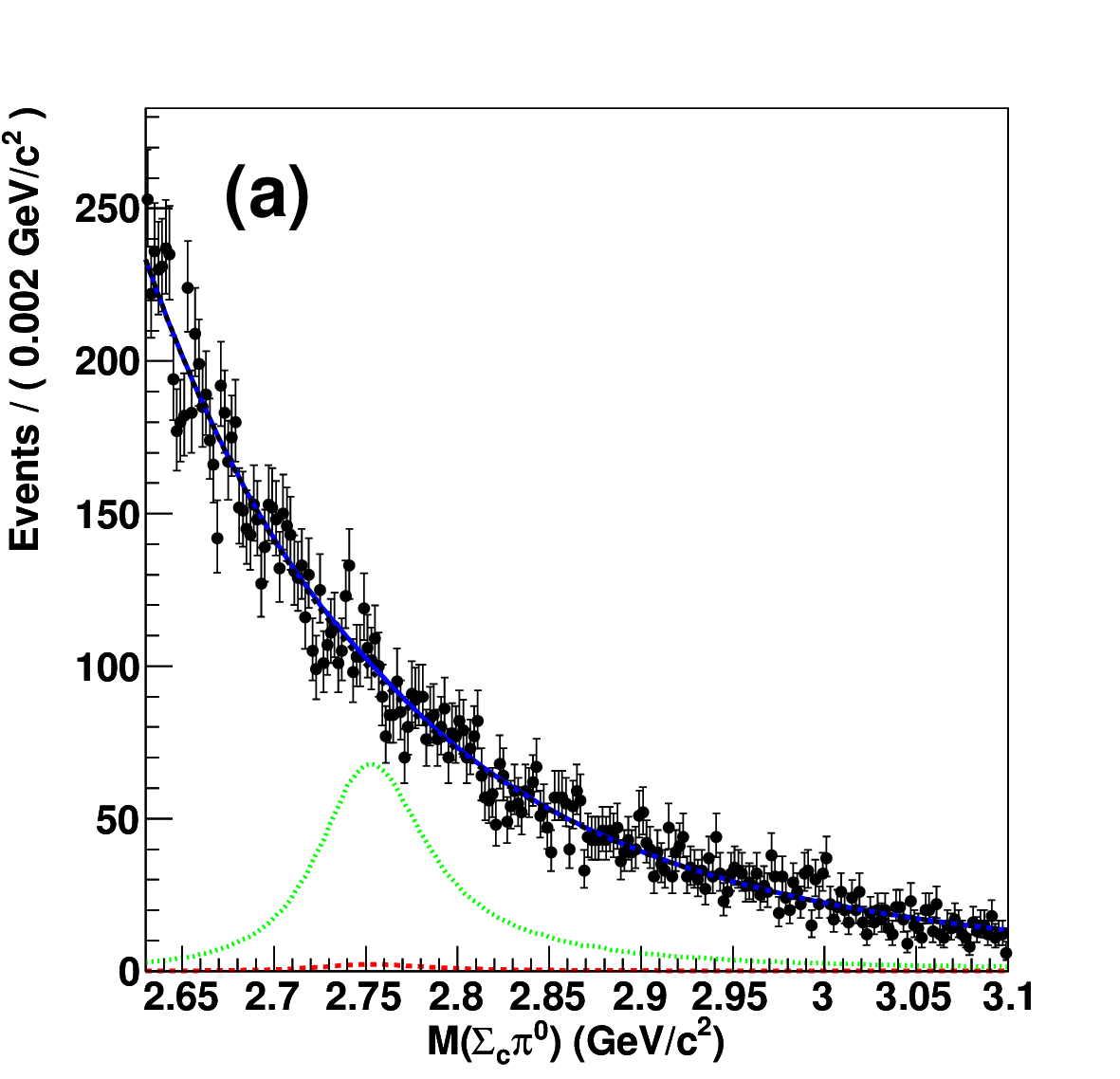}
    \includegraphics[scale=0.35]{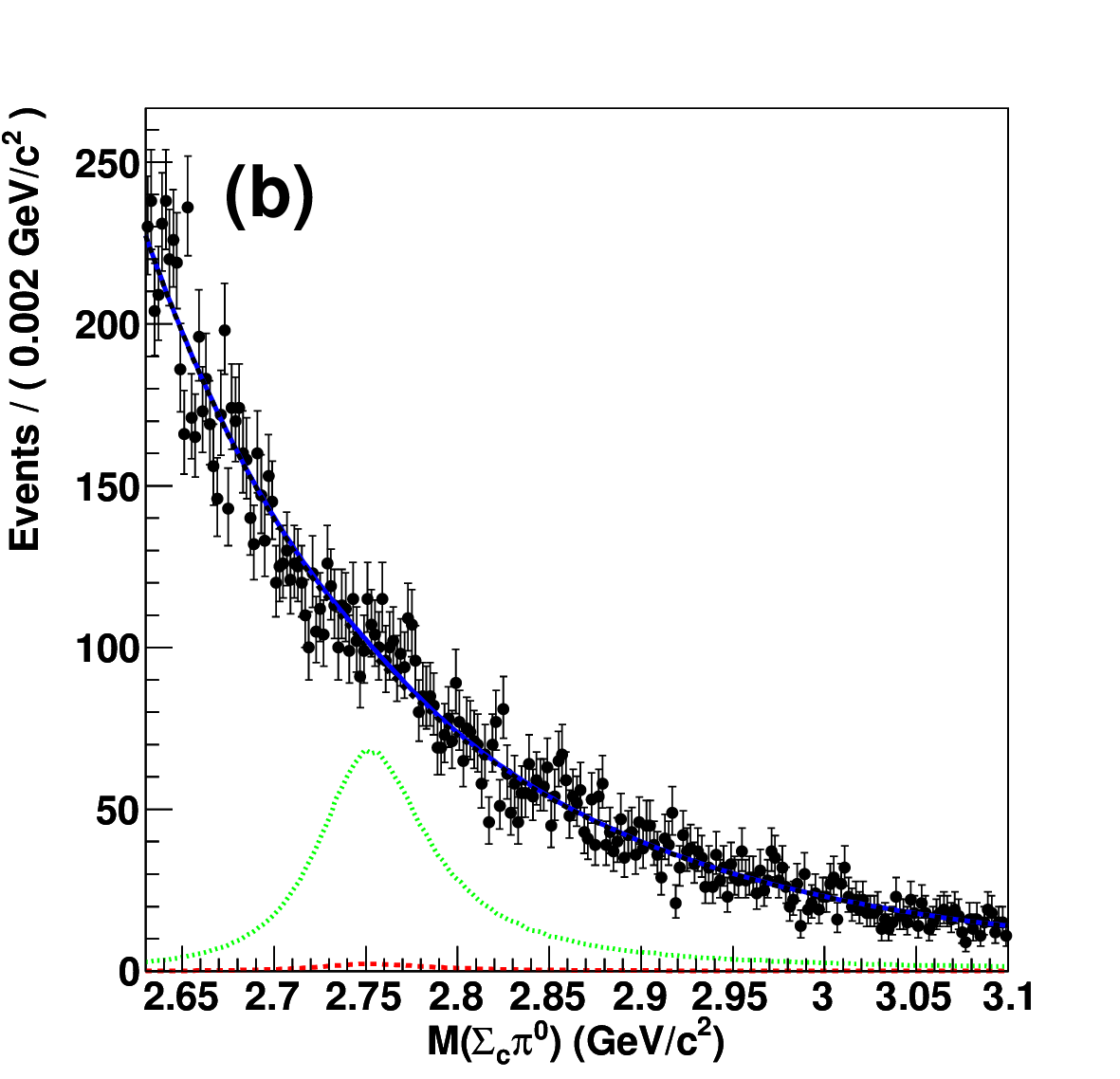}
    \caption{ (a) $M(\Lambda_{c}^{+}\pi^{+}\pi^{0})$ and (b) $M(\Lambda_{c}^{+}\pi^{-}\pi^{0})$ distributions
	      with $\Sigma_{c}(2455)^{++/0}$ selection for data with the fit results (blue solid lines).
             The red dashed lines are the signal contributions from the fit, while
             the green dotted lines show the expected signals in case the production cross section 
             and branching fractions are the same as $\Sc(2765)^+$.}
    \label{box_opening_sc2455}
  \end{center}
\end{figure*}

 Fig.~\ref{box_opening_sc2455} shows 
$M(\Lambda_{c}^{+} \pi^{\pm} \pi^{0})$ distributions for 
the signal regions with fit lines ovarlaid.
A simultaneous unbinned extended maximum likelihood fit 
is performed with the constraint that the signal yields for 
$I_3=\pm1$ states are the same.
The signal PDF is made based on the signal MC simulation assuming 
that the mass and width are the same as those of the $I_{3}=0$ state.
The sum of the signal yields for $I_{3}=\pm1$ is obtained to be $230\pm250$. 
Note that, if we assume that the cross-section and 
branching fractions are the same, the expected
signal yields would be 8764, estimated from the yields of
$\Lcts \to \Sc(2455)^{++/0}\pi^{\mp} \to \Lc \pi^+ \pi^-$
taking the difference of acceptance and efficiency into account.

The $\chi^{2}/d.o.f.$ of the fit is calculated 
after filling the data in bins as
211/232 (215/232) for $M(\Lambda_{c}^{+} \pi^{+} \pi^{0})$ 
[$M(\Lambda_{c}^{+} \pi^{-} \pi^{0})$]
without including the signal component.
The 95$\%$ upper limit on the $R_{I}$ is obtained based 
on Bayesian approach, after incorporating the following systematic uncertainties
which are summarized in Table \ref{summary_sys_isospin}.
The obtained 95$\%$ C.L upper limit for $R_{I}$ is 0.09, which is much smaller than unity,
demonstrating that isospin of $\Lambda_{c}(2765)^{+}$ is indeed zero.
This is consistent with the fact that the state was not seen in 
the $\Lc\pi$ spectra by Belle \cite{Belle-Sc2800}, that would be an 
expected decay mode of an excited $\Sigma_c$ baryon.

\begin{table*}[htbp]
  \begin{center}
    \caption{Summary of the systematic uncertainties for $R_{I}$.}
    \begin{tabular}{c|c} \hline \hline
      Source                                       &  ($\%$)      \\ \hline
      $\pi^{0}$   reconstruction efficiency        &2.1           \\ 
      $\pi^{+/-}$ reconstruction efficiency        &0.9           \\ 
      $\pi^{+/-}$ Track reconstruction  efficiency &0.4           \\ 
      Fit bias                                     &0.3           \\ 
      Yield for $I_{3}=0$ state                    &25.3          \\ 
      Width for $I_{3}=0$ state                    &17.1          \\ 
      Mass of   $I_{3}=0$ state                    &1.7           \\ \hline \hline
      Total                                        &30.7          \\
    \end{tabular}
    \label{summary_sys_isospin}
  \end{center}
\end{table*}

\section{Summary and discussion}
\label{discussion}

In summary, we determined the isospin of 
$\Lcts/\Sigma_c(2765)^+$ by searching
for possible isospin partners in the decays to 
$\Sc(2455)^{++/0}\pi^0 \to \Lc \pi^{\pm} \pi^0$.
The isospin is determined to be $I=0$, 
therefore the name should indeed be $\Lcts$.
The result is rather inconsistent with the most recent 
quark model calculation \cite{th-yoshida}, but not with 
the older ones \cite{th_capstick, th-Roberts}. 
Also, some theoretical predictions are ruled out by
the result. However, in order to discuss the 
nature of this state in more detail,
the determination of its other quantum numbers 
(i.e., spin and parity) is highly desirable and we
are planning a publication on this subject, 
using this data set, in the near future.

\section*{Acknowledgment}

We thank the KEKB group for the excellent operation of the
accelerator; the KEK cryogenics group for the efficient
operation of the solenoid; and the KEK computer group, and the Pacific Northwest National
Laboratory (PNNL) Environmental Molecular Sciences Laboratory (EMSL)
computing group for strong computing support; and the National
Institute of Informatics, and Science Information NETwork 5 (SINET5) for
valuable network support.  We acknowledge support from
the Ministry of Education, Culture, Sports, Science, and
Technology (MEXT) of Japan, the Japan Society for the 
Promotion of Science (JSPS) under KAKENHI Grant No. JP19H05148, 
and the Tau-Lepton Physics 
Research Center of Nagoya University; 
the Australian Research Council including grants
DP180102629, 
DP170102389, 
DP170102204, 
DP150103061, 
FT130100303; 
Austrian Science Fund (FWF);
the National Natural Science Foundation of China under Contracts
No.~11435013,  
No.~11475187,  
No.~11521505,  
No.~11575017,  
No.~11675166,  
No.~11705209;  
Key Research Program of Frontier Sciences, Chinese Academy of Sciences (CAS), Grant No.~QYZDJ-SSW-SLH011; 
the  CAS Center for Excellence in Particle Physics (CCEPP); 
the Shanghai Pujiang Program under Grant No.~18PJ1401000;  
the Ministry of Education, Youth and Sports of the Czech
Republic under Contract No.~LTT17020;
the Carl Zeiss Foundation, the Deutsche Forschungsgemeinschaft, the
Excellence Cluster Universe, and the VolkswagenStiftung;
the Department of Science and Technology of India; 
the Istituto Nazionale di Fisica Nucleare of Italy; 
National Research Foundation (NRF) of Korea Grants
No.~2015H1A2A1033649, No.~2016R1D1A1B01010135, No.~2016K1A3A7A09005603,
No.~2016R1D1A1B02012900, No.~2018R1A2B3003643,
No.~2018R1A6A1A06024970, No.~2018R1D1 A1B07047294; Radiation Science Research Institute, Foreign Large-size Research Facility Application Supporting project, the Global Science Experimental Data Hub Center of the Korea Institute of Science and Technology Information and KREONET/GLORIAD;
the Polish Ministry of Science and Higher Education and 
the National Science Center;
the Grant of the Russian Federation Government, Agreement No.~14.W03.31.0026; 
the Slovenian Research Agency;
Ikerbasque, Basque Foundation for Science, Spain;
the Swiss National Science Foundation; 
the Ministry of Education and the Ministry of Science and Technology of Taiwan;
and the United States Department of Energy and the National Science Foundation.


\begin{thebibliography}{99}
\bibitem{footnote-notation} 
In this paper, the notation $\Sigma_c(2765)$ is used only
in the discussions of isospin partners of $\Sigma_c(2765)^+$ and
$\Lcts$ or $\Lcts/\Sigma_c(2765)^+$ is used in the other cases. 
The choice of $\Lcts$ is conventional and is consistent with the
conclusion of the paper.
\end{thebibliography}
\end{document}